\documentclass[fleqn,twoside]{article}
\usepackage{espcrc2}
\usepackage{graphicx}
\usepackage[figuresright]{rotating}
\title{ $\tau^-\to K^+K^-\pi^-\nu_\tau$  at CLEO III } 
\author{Feng Liu 
\address{On behalf of  the CLEO Collaboration\\ 
SMU Group, Wilson Lab, Cornell University, Ithaca, NY 14853, USA }  
\thanks{E-mail: feng@mail.lns.cornell.edu} } 
\begin{document}

\begin{abstract}
Based on 3.26 fb$^{-1}$ data sample collected at CLEO III, we study the 
decay $\tau^-\to K^+K^-\pi^-\nu_\tau$, and improve 
the measurement of the branching ratio 
${\cal B}(\tau^-\to K^+K^-\pi^-\nu_\tau)=(1.59\pm0.06\pm0.13)\times10^{-3}$, 
the results are preliminary. 
\end{abstract}
\maketitle
 
\section{Introduction} 

For $\tau$ hadronic decays, final states
with kaons provide a powerful probe of the strange sector of the 
weak charged current. The decay channel 
$\tau^-\to K^+K^-\pi^-\nu_\tau$, together with
$\tau^-\to\eta\pi^-\pi^0\nu_\tau$ and $\tau^-\to K^-\pi^+\pi^-\nu_\tau$  
can be used~\cite{prd47} to  measure %quantitatively determine 
the contribution %to $\tau$ decays into three mesons 
from the Wess-Zumino anomaly~\cite{WZ}.  
Furthermore,  $\tau^-\to K^+K^-\pi^-\nu_\tau$ with sizable branching ratio 
and higher hadronic mass, can be used  to give more stringent
constraints on the $\tau$ neutrino mass~\cite{BaBar}. 

Table~\ref{brlist} lists the recent  measurements of 
$\tau^-\to K^+K^-\pi^-\nu_\tau$ branching fraction  
~\cite{br4}. These measurements are based on the particle separation provided 
by $dE/dx$. There exists the discrepancy between the 
result from OPAL and those from the other two.
% (OPAL, ALEPH and CLEO II, DELCO used Cherenkov counter). 
At CLEO III, we have about 9.2 fb$^{-1}$ data (full dateset)
 taken at $\sqrt s\sim 10.58$ GeV, out of which 
3.26 fb$^{-1}$ (corresponding to 2.97$\times10^6$ $\tau^+\tau^-$ pairs) 
is available for this analysis.
 The RICH detector at CLEO III provides good particle separation, 
allowing us to improve the measurement of $\tau^-\to K^+K^-\pi^-\nu_\tau$
branching fraction. 
\begin{table}[h] 
%\vskip -0.5cm 
\caption{Measurements of ${\cal B}(\tau^-\to K^+K^-\pi^-\nu_\tau)$.} 
\begin{center} 
\begin{tabular}{|c|c|c|}\hline 
Group & $\tau^+\tau^-(\times10^6)$  & ${\cal B} (\%)$ \\ \hline 
OPAL  & 0.15 & $0.087\pm0.056\pm0.040$ \\ \hline 
ALEPH & 0.15 & $0.163\pm0.021\pm0.017$ \\ \hline
CLEO II & 4.3  & $0.145\pm0.013\pm0.028$ \\ \hline 
%DELCO &  0.03  & $0.22^{+0.17}_{-0.11}\pm0.05$ \\ \hline
\end{tabular}
\end{center} 
\label{brlist} 
\end{table} 

\section{Event Selection} 

We use 1-prong $\tau$ leptonic decay ($e/\mu$) or hadronic 
decay ($\rho/\pi(K)$) to tag the other $\tau^-\to K^+K^-\pi^-\nu_\tau$.  
We select the events with the 1 vs. 3 topology. And a event 
is required to have 4 tracks, passing the track quality cuts with 
net zero charge.   
An electron candidate is selected based on $dE/dx$ information and the ratio of the 
associated shower energy in the calorimeter to the measured track momentum. 
A muon candidate must penetrate 
at least  three  or five absorption  lengths of material depending on its momentum. 
Any hadron candidate is required not to be identified as an electron or a muon.

The events that do not satisfy the lepton criteria are classified as 
hadron tags. To find a $\rho$, we use the most energetic
$\pi^0$ candidate in the tagging hemisphere within 
$(-3.0,2.5)\sigma$ of the $\pi^0$ nominal mass, 
the $\pi^0$ candidate
with a charged pion in the tagging hemisphere must be consistent with the $\rho$
 mass. { All other tags with an invariant mass in the 
tagging hemisphere less than 0.5 GeV are classified as $\pi
$.} 
To select the photon candidates from $\pi^0$ decays, 
unmatched showers with $|\cos\theta|<0.95$ are required 
to have energies greater than 30 MeV in the good section of the 
 barrel or 50 MeV in the endcap. To get better mass resolution, 
we require that 
at least one photon candidate from $\pi^0$ decays must be in the good 
section of the barrel. 

A typical signal event should not deposit any significant 
extra energy in the calorimeter. In most cases, extra
unmatched energy deposited in the calorimeter is a signature
of various backgrounds with one or more $\pi^0$'s. To 
suppress these backgrounds,  
we don't allow %the events are not  allowed to have  
any extra unmatched shower with energy $E_\gamma>0.1$ GeV
in the tagging hemisphere or in the signal hemisphere. 
 
%At this stage, there are considerable two photon backgrounds remaining 
%in the data. They are characterized by the 
%missing momentum along the beam pipe and low visible energy. 
In order  to reject the two photon background which is characterized by the 
missing momentum along the beam pipe and a small visible energy, 
we require $|\cos\theta_{miss}|<0.95$, $E_{vis}/E_{cm}>0.4$.
For $\tau^-\to K^+K^-\pi^-\nu_\tau$ decay with $\nu_\tau$ undetected, the 
$K^+K^-\pi^-$ system has high transverse momentum,  
%In order to further suppress continuum and two photon backgrounds, 
we require $p_t^{KK\pi}>2.0$ GeV/c

To get better particle separation, $\sigma_{\pi,K}$ (the standard 
deviation of the measured $dE/dx$ from the expected one)  
and RICH loglikelihood difference $\chi^2_{\pi}-\chi^2_{K}$ 
under pion and kaon hypotheses were
combined together. The particle identification (PID) cuts 
$\Delta \chi^2=\chi^2_{\pi}-\chi^2_{K}+
\sigma_\pi^2-\sigma_K^2$ and the number of 
Cherenkov photons $N_\gamma^K$ are evaluated  based on 
generic $\tau$ MC and  generic continuum MC,  and  
 are determined as     
$N_\gamma^K\ge3$, $\Delta \chi^2>0$ for kaons;
and $|\sigma_\pi|<3.0$ for pion.  

\subsection{The Backgrounds} 

After the event selection,
 two photon backgrounds are totally negligible.
The main backgrounds
come from $\tau$ decays to three charged tracks with or without neutrals.  
%(${\cal B}(\tau^-\to\pi^+\pi^-\pi^-\ge0$ neutrals)=14.5\% v.s 
%${\cal B}(\tau^-\to K^+K^-\pi^-\nu_\tau)=0.161\%$~\cite{PDG} ). 
We use the generic continuum MC and generic $\tau$ MC to study the 
continuum and $\tau$  backgrounds.  
With RICH, the continuum backgrounds are suppressed to about 
2\% level, and the $\tau$ backgrounds are also well suppressed to about 9\% 
level.

\section{Results} 
 After the event selection, 
we observe 
938 events, out of which, 19 and 86 events are from 
the continuum and $\tau$ backgrounds, respectively. 
After the backrounds subtraction, we observe $833\pm32$ 
signal events, the overall efficiency is (12.28$\pm$0.11)\%. 
From those, we determine the branching ratio 
${\cal B}(\tau^-\to K^+K^-\pi^-\nu_\tau)=(1.59\pm0.06({stat}))\times10^{-3}$.   

\begin{figure}[tb]
\vspace{9pt}
%\framebox[55mm]{\rule[-21mm]{0mm}{43mm}}
\includegraphics[width=0.5\textwidth]{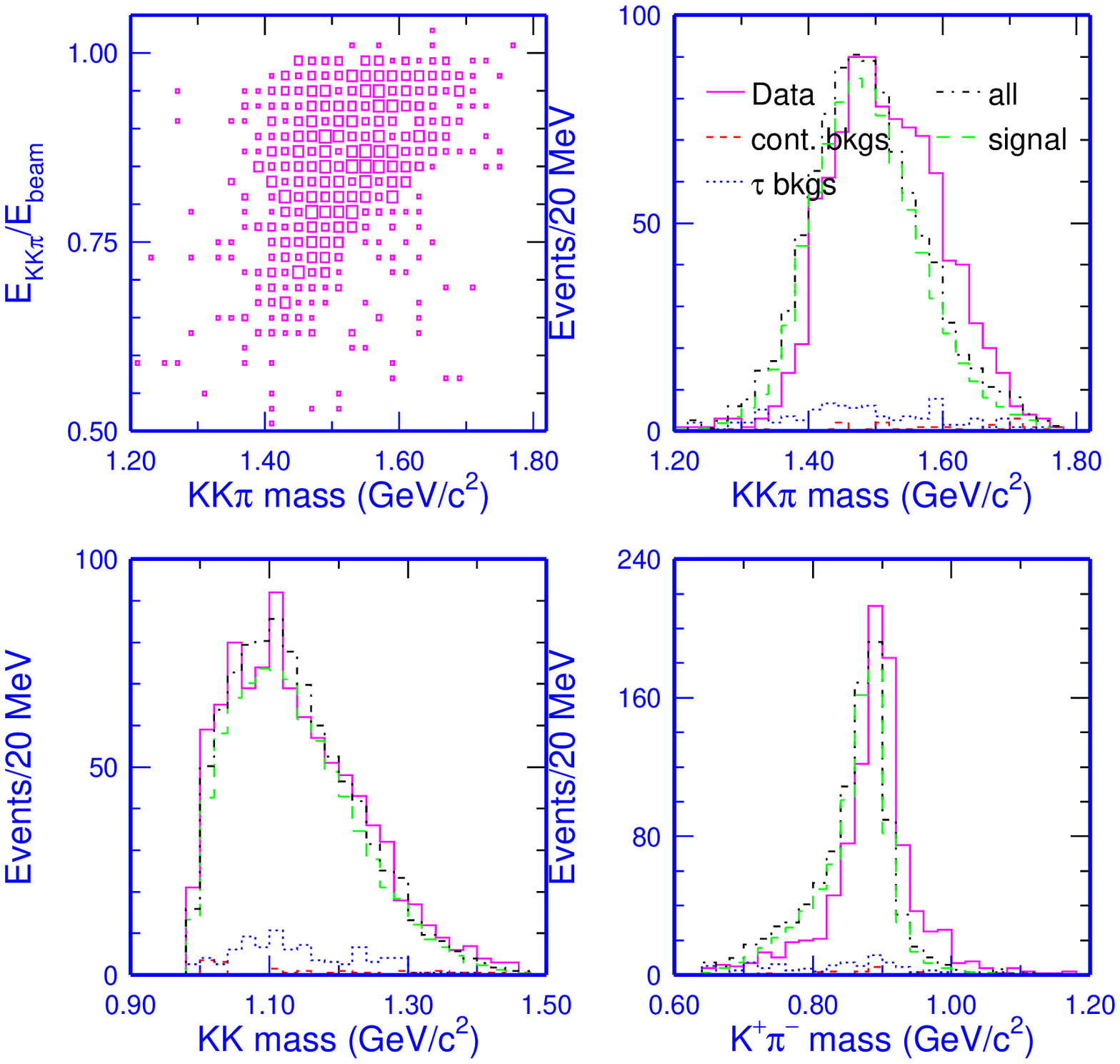}
\caption{$\tau\to K^+K^-\pi^-\nu_\tau$ mass spectra.}  
\label{mass} 
\end{figure}

The  substructure of $\tau^-\to K^+K^-\pi^-\nu_\tau$ decay 
provides a way to measure the 
Wess-Zomino anomaly~\cite{prd47}. In Fig.~\ref{mass}, we  show the mass
spectra of hadronic systems. It is clear that $\tau\to K^+K^-\pi^-\nu_\tau$ 
is dominated by the intermediate state $\tau^-\to K^-K^{*0}\nu_\tau$ with
$K^{*0}\to K^+\pi^-$, there is 
no evidence for the $\phi$ production. The nomalized continuum and $\tau$ 
backgrounds are overlaid. The comparison between the data and MC 
(all=backgrounds+signal) %is also presented. The comparison 
shows that the decay is not 
well modelled in korb~\cite{korb}. The modelling 
 of the substructure needs improvement.
 
\section{Systematics} 
\label{systematics4}
The systematics mainly comes from the event selection efficiencies. 
To get the systematics, we vary the cuts by 10\% (or more) or 1$\sigma$ 
at the determined values and calculate 
the new yields, efficiencies and branching ratios to obtain the systematics.
The systematics from the $E_\gamma<0.10$ GeV requirement for
the extra shower is less than 1\%. 
 Since $\tau^-\to K^+K^-\pi^-\nu_\tau$
is not well modeled in Korb~\cite{korb}, we use phase 
space generator to generate $\tau^-\to K^+K^-\pi^-\nu_\tau$ and 
$\tau^-\to K^-K^{*0}\nu_\tau$ where $K^{*0}\to K^+\pi^-$, 
and find the systematics from 
the model dependent efficiency to be 1.5\% and 1.8\%, respectively.  The 
systematics %from the continuum background subtraction and 
from the $\tau$ background subtraction due to the uncertainties 
in the branching ratios for $\tau$ three prong decays
is 1.3\%. The systematics from the continuum background subtraction is
estimated to be 1.5\%.  The systematics from the cut on the number of tracks
in a event is 1.7\%. 
 The tracking efficiency systematics is 0.2\% per track, the PID efficiency
systematics is 3.0\% per kaon and 1.0\% per pion.  Systematics also includes 
the uncertainties from the $\tau$ pair cross section (1\%), the integrated 
luminosity (2\%) and MC statistics (1\%). The 
total systematics is 8.1\%, the dominant contribution comes 
from the PID efficiency systematics, and we have 
${\cal B}(\tau^-\to K^+K^-\pi^-\nu_\tau)=(1.59\pm0.06(stat)\pm0.13(sys))\times10^{-3}$ (preliminary).

\section{Cross Check: Measurement of ${\cal B}(\tau^-\to\pi^+\pi^-\pi^-\nu_\tau$) } 
\begin{figure}[htbp]
\vspace{9pt}
%\framebox[55mm]{\rule[-21mm]{0mm}{43mm}}
\includegraphics[width=0.5\textwidth]{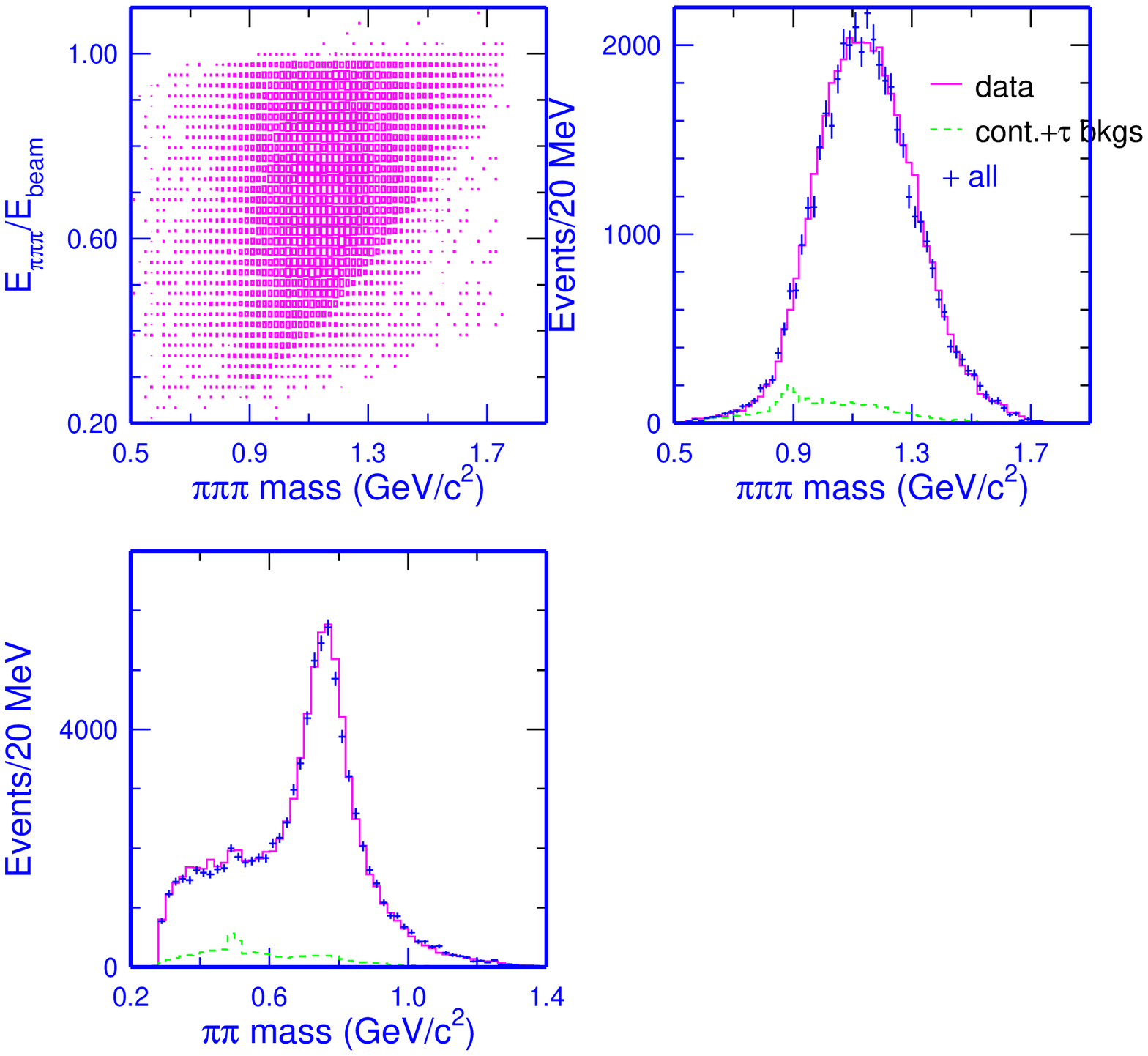}
\caption{$\tau\to \pi^+\pi^-\pi^-\nu_\tau$ mass spectra.}  
\label{com3pi} 
\end{figure}

As a cross check, we present the measurement of ${\cal B}(\tau^-\to\pi^+\pi^-\pi^-\nu_\tau$). 
We use $N_\gamma^\pi\ge3$ and
 $\Delta \chi^2=\chi^2_{\pi}-\chi^2_{K}+ \sigma_\pi^2-\sigma_K^2<0$ to
identify three pions.
After the event selection, we observe 43,432 events, the continuum background is
151.6 events, the $\tau$ background is 3,207 events. 
After the backgrounds subtracted,  
we observe 40,073$\pm$216 signal events. Fig.~\ref{com3pi} 
show the comparison of the mass sepectra between the data and MC. 
The agreement is quite good.  The overall efficiency 
is (10.21$\pm$0.17)\%, and we have 
${\cal B}(\tau^-\to \pi^+\pi^-\pi^-\nu_\tau)=(9.21\pm0.05)\%$. 
The pion PID efficiency  systematics is 1\% per pion, MC statistics is 2\%, the 
total systematics is about 5\% (reference to Sec.~\ref{systematics4}). Thus we obtain 
${\cal B}(\tau^-\to \pi^+\pi^-\pi^-\nu_\tau)=(9.21\pm0.05\pm0.46)\%$. The result
is in good agreement with the PDG value 
${\cal B}(\tau^-\to \pi^+\pi^-\pi^-\nu_\tau)=(9.18\pm0.11)\%$.  

\section{Summary} 
Based on 3.26 fb$^{-1}$ ($2.97\times10^6$ $\tau^+\tau^-$)
data collected at CLEO III, we improve the measurement of 
${\cal B}(\tau^-\to K^+K^-\pi^-\nu_\tau)=(1.59\pm0.06\pm0.13)\times10^{-3}$  
which is in good agreement with the PDG result 
$(1.61\pm0.18)\times10^{-3}$~\cite{PDG}, the errors are statistical 
and systematic, respectively. 
 The systematics is dominated by the 
PID efficiency systematics which we'll further
 study after more data available. The results are preliminary. 
The modelling of the substructure is under study. 

%\section*{Acknowledgments} 
%We  like to thank Alan Weinstein and Jon Urheim 
%and our paper committee members: Brian K. Heltsley (chair) and 
%Bayar Dambasuren for their very helpful comments and suggestions.  

\end{document}